\documentclass[aps,prd,twocolumn,nofootinbib,superscriptaddress,floatfix]{revtex4-1}
\usepackage{graphicx}
\usepackage{bm}
\usepackage{amsmath}
\usepackage{amssymb}
\usepackage{amsfonts}
\usepackage{hyperref}
\hypersetup{colorlinks=true,citecolor=blue,linkcolor=blue,urlcolor=blue}
\usepackage{tabularx}
\usepackage[normalem]{ulem}
\usepackage{color}

\newcommand*\aap{A\&A}

\newcommand*\jcap{J. Cosmology Astropart. Phys.}
\newcommand*\mnras{MNRAS}

\defcitealias{PPG.etal.2018}{P18}

\begin{document}


\title{Distinguishing standard and modified gravity cosmologies with machine learning}

\author{Austin Peel}\email{austin.peel@cea.fr}
\affiliation{AIM, CEA, CNRS, Universit{\'e} Paris-Saclay, Universit{\'e} Paris Diderot, 
             Sorbonne Paris Cit{\'e}, F-91191 Gif-sur-Yvette, France}
\author{Florian Lalande}
\affiliation{AIM, CEA, CNRS, Universit{\'e} Paris-Saclay, Universit{\'e} Paris Diderot, 
             Sorbonne Paris Cit{\'e}, F-91191 Gif-sur-Yvette, France}
\affiliation{ENSAI, rue Blaise Pascal, 35170 Bruz, France}
\author{Jean-Luc Starck}
\affiliation{AIM, CEA, CNRS, Universit{\'e} Paris-Saclay, Universit{\'e} Paris Diderot, 
             Sorbonne Paris Cit{\'e}, F-91191 Gif-sur-Yvette, France}
\author{Valeria Pettorino}
\affiliation{AIM, CEA, CNRS, Universit{\'e} Paris-Saclay, Universit{\'e} Paris Diderot, 
             Sorbonne Paris Cit{\'e}, F-91191 Gif-sur-Yvette, France}
\author{\\Julian Merten}
\affiliation{INAF -- Osservatorio di Astrofisica e Scienza dello Spazio di Bologna,
             via Gobetti 93/3, 40129, Bologna, Italy}
\author{Carlo Giocoli}
\affiliation{INAF -- Osservatorio di Astrofisica e Scienza dello Spazio di Bologna,
             via Gobetti 93/3, 40129, Bologna, Italy}
\affiliation{Dipartimento di Fisica e Astronomia, Alma Mater 
             Studiorum Universit\`{a} di Bologna, via Gobetti 
             93/2, 40129 Bologna, Italy}
\affiliation{Dipartimento di Fisica e Scienze della Terra, 
             Universit\`{a} degli Studi di Ferrara, via Saragat 1, 
             I-44122 Ferrara, Italy}
\affiliation{INFN -- Sezione di Bologna, viale Berti Pichat 6/2, 
             40127, Bologna, Italy}
\author{Massimo Meneghetti}
\affiliation{INAF -- Osservatorio di Astrofisica e Scienza dello Spazio di Bologna,
             via Gobetti 93/3, 40129, Bologna, Italy}
\affiliation{Dipartimento di Fisica e Astronomia, Alma Mater 
             Studiorum Universit\`{a} di Bologna, via Gobetti 
             93/2, 40129 Bologna, Italy}
\affiliation{INFN -- Sezione di Bologna, viale Berti Pichat 6/2, 
             40127, Bologna, Italy}
\author{Marco Baldi}
\affiliation{INAF -- Osservatorio di Astrofisica e Scienza dello Spazio di Bologna,
             via Gobetti 93/3, 40129, Bologna, Italy}
\affiliation{Dipartimento di Fisica e Astronomia, Alma Mater 
             Studiorum Universit\`{a} di Bologna, via Gobetti 
             93/2, 40129 Bologna, Italy}
\affiliation{INFN -- Sezione di Bologna, viale Berti Pichat 6/2, 
             40127, Bologna, Italy}
\date{\today}

\begin{abstract}
We present a convolutional neural network to classify distinct cosmological scenarios based on the statistically similar weak-lensing maps they generate. Modified gravity (MG) models that include massive neutrinos can mimic the standard concordance model ($\Lambda$CDM) in terms of Gaussian weak-lensing observables. An inability to distinguish viable models that are based on different physics potentially limits a deeper understanding of the fundamental nature of cosmic acceleration. For a fixed redshift of sources, we demonstrate that a machine learning network trained on simulated convergence maps can discriminate between such models better than conventional higher-order statistics. Results improve further when multiple source redshifts are combined. To accelerate training, we implement a novel data compression strategy that incorporates our prior knowledge of the morphology of typical convergence map features. Our method fully distinguishes $\Lambda$CDM from its most similar MG model on noise-free data, and it correctly identifies among the MG models with at least 80\% accuracy when using the full redshift information. Adding noise lowers the correct classification rate of all models, but the neural network still significantly outperforms the peak statistics used in a previous analysis.
\end{abstract}



\maketitle

\section{Introduction}\label{sec:intro}
The quest to uncover the physical origin of cosmic acceleration remains one of the biggest challenges in modern cosmology. Whether the source is a cosmological constant, a dynamical fluid, or a modification of gravity is not yet known. Although cosmic microwave background analyses are compatible with a Lambda cold dark matter ($\Lambda$CDM) model \cite{Planck.DE.2016, Planck.Cosmo.2018}, degeneracies exist between standard parameters and modified gravity models in terms of second-order statistics, particularly in the presence of massive neutrinos \cite[e.g.][]{MSY.2013,He.2013,BVNV.etal.2014,WWK.2017}. The theoretical challenge is then coupled to a data analysis challenge: determining whether more advanced statistics or still new methods can help break such degeneracies.

Weak gravitational lensing (WL) is now firmly established as a robust observational probe for cosmological inference. Galaxy shear measurements have been used to test various cosmological models in both general relativistic (i.e. $\Lambda$CDM) and nonstandard (e.g. modified gravity, MG) frameworks \cite[e.g.][]{TT.2008, Schmidt.2008, BBK.2010, DLS.etal.2010, TSS.2011, CDC.2011, CFHTLenS.WL.2013, LBF.2015, KIDS.WL.2017, DES.WL.2017}. Despite the successes of weak lensing, the question remains open of how to best extract and leverage the information contained in galaxy surveys that provide these data.

In previous work \cite[][hereafter P18]{PPG.etal.2018}, we proposed higher-order statistics in the weak-lensing signal as a new set of observables capable of breaking degeneracies between massive neutrinos and MG parameters. In particular, we showed that peak count statistics outperform higher-order moments like skewness and kurtosis in distinguishing between $\Lambda$CDM and $f(R)$ models when the latter contain massive neutrinos. We found that the discrimination efficiency depends, however, on redshift and the observed angular scale. Furthermore, including galaxy shape noise degrades the discrimination power across all aperture scales, source redshifts, and statistics, including peak counts. While a de-noising procedure may help, we present here a new method that provides significantly better results on both clean and noisy maps.

Recent advances in the field of machine learning (ML) have led to new analysis tools for cosmology. Such tools can be used to break the well-known degeneracy between the standard model parameters $\sigma_8$ and $\Omega_\mathrm{m}$ that arises in weak lensing. For example, a Convolutional Neural Network (CNN) trained on (noisy) convergence maps was shown to discriminate models along this degeneracy better than skewness and kurtosis \cite{SLK.etal.2017}. CNNs have also obtained tighter constraints on the $\sigma_8$--$\Omega_\mathrm{m}$ confidence contour than traditional Gaussian (e.g. power spectrum) and non-Gaussian (e.g. peak statistics) observables \cite{GMH.etal.2018, FKL.etal.2018}. An interpretation of the filters learned by such networks points to the steepness (as opposed to just the amplitude) of local peaks as carrying the most salient cosmological information \cite{RAC.2018}. Beyond the weak-lensing domain, ML algorithms have also been applied to investigate large-scale structure formation \cite{LSPP.etal.2018, BS.2018} and the connection between baryonic and dark matter distributions \cite{KTB.2016, ADB.2018, NMW.etal.2018}.

In light of results from \citetalias{PPG.etal.2018}, we aim here to determine whether a ML neural network can discriminate better than peak counts between $\Lambda$CDM and MG models with massive neutrinos. We study the same weak-lensing maps as in \citetalias{PPG.etal.2018}, which are derived from a subset of the {\small DUSTGRAIN}-{\em pathfinder} simulations \cite{GBM.2018}. These cosmological models represent viable locations in the $f_{R0}$--$M_\nu$ parameter space that are intentionally difficult to distinguish from $\Lambda$CDM via only second-order statistics. The simulations have been used separately to forecast cosmological constraints from measurements of the halo mass function of dark matter \cite{HCB.etal.2018}.

We consider one $\Lambda$CDM model without neutrinos and three $f(R)$ modified gravity models with different neutrino mass sums $M_\nu := \sum m_\nu\in\{0, 0.1, 0.15\}~\mathrm{eV}$, where we fix the parameters $n=1$ and $f_{R0}=-10^{-5}$ in the Hu-Sawicki formulation \cite{HS.2007}. As was shown in \citetalias{PPG.etal.2018}, the $\Lambda$CDM model is most degenerate with the $f(R)$ model with $M_\nu=0.15~\mathrm{eV}$ when relying only on second-order statistics. By degenerate, we mean that their noise-free (average) convergence power spectra agree at better than 10\% over the full range of angular scales and source redshifts considered. In practice, these two models, as well as other neighboring pairs of $f(R)$ models, agree at better than 5\% for all but a small range of $\ell$ modes. Accounting for galaxy shape noise, as we explore in this work, further increases the similarity of predicted second-order observables. Improving the discrimination efficiency between pairs of such similar models is the main goal of our proposed methodology.

The remainder of the paper is organized as follows. In Sec.~\ref{sec:methodology} we introduce our ML methodology, including an approach to accelerate training by reducing the dimensionality of the problem. In Sec.~\ref{sec:performance} we describe measures of monitoring the progress of training and our strategy to evaluate the network's performance on test data. We present our results in Sec.~\ref{sec:results} on the correct classification rate of clean and noisy convergence maps, and we discuss our conclusions in Sec.~\ref{sec:conclusions}.

\section{Methodology}\label{sec:methodology}
Given the parameters of the {\small DUSTGRAIN}-{\em pathfinder} simulations, we obtained 256 pseudo-independent convergence maps for each model using the \textsc{MapSim} pipeline \cite{GMB.etal.2015}. Because the past light cones were produced by randomly reorienting the same particle field in a given simulation run, different lines of sight are not strictly independent if their light cones intersect somewhere. A large box size, however, means that the probability of such intersections is small, and we thus do not expect this approximation to affect our analysis.

Each derived convergence map covers a $25~\mathrm{deg}^2$ square area of sky sampled at a resolution of $2048 \times 2048$ pixels. Six maps were computed for each line of sight assuming source galaxy redshift planes at $z_s \in \{0.5, 1.0, 1.5, 2.0, 3.0, 4.0 \}$. In this work we consider only the four redshifts up to $z_s=2.0$ in order to compare with previous results \citepalias{PPG.etal.2018} and because this is roughly the range accessible to current and planned weak-lensing surveys.

Using such large maps as direct inputs to a CNN can represent a training challenge given limited computing resources. We propose to alleviate this issue by reducing the dimensionality of our data before training, that is, by condensing each map's information content. To do this effectively, we incorporate our prior expectation of the morphology of typical features present in weak-lensing maps in order to change their representation without losing information relevant to the problem. Our approach uses the probability distribution function (PDF) of map coefficients at each scale of a multi-scale wavelet transform. The specific choice of wavelet function is a good match to the local peak structures in weak-lensing maps.

\subsection{Wavelet PDF representation}\label{subsec:wavelets}
Wavelet transforms have found numerous applications in astronomical image processing. In particular, the starlet (isotropic undecimated wavelet) transform provides a useful representation space for weak-lensing convergence maps \cite{SPR.2006, PSA.etal.2009, LPS.2012, PLS.2012, LLS.2014, LSL.etal.2016, LKP.2016, PLL.etal.2017}. This transform naturally facilitates a multi-scale analysis: an initial $N \times N$ map is decomposed into $j_{\max}$ wavelet coefficient maps labeled as $w_j$, where $j \in \{1,...,j_{\max}\}$, plus a final coarse-scale map. The coarse-scale map is a low-pass filtered version of the original image. Each $w_j$ (also $N \times N$), on the other hand, represents a band-pass filtering and is equivalent to an aperture mass map filtered at a scale of $2^j$ pixels \cite{LPS.2012}.

Recent results \cite{RAC.2018} also motivate the use of wavelets in machine learning applications to weak-lensing analysis. In this work, the authors found that the first hidden layer of their CNN, trained on convergence maps, learned filters extracting information about the gradients around peaks. The compensated shape of our wavelet functions extracts precisely this type of information as well and at multiple scales simultaneously. We therefore consider our pre-processing with wavelets as a time-saving step that avoids the network having to learn this first layer, thereby reducing the number of trainable parameters as well as the amount of training data necessary.

Given an initial convergence map, we derive its condensed wavelet PDF representation as follows: (1) compute the starlet transform with $j_{\max}=5$, corresponding to a minimum (maximum) aperture smoothing of 0.293 (4.69) arcmin; (2) divide each $w_j$ map by the standard deviation of all 1024 maps (256 per model) at the same redshift and scale; (3) bin the resulting pixel values into arrays with 100 steps between $-5.0$ and $5.0$ per scale and normalize by the integral over the range to obtain the PDF; (4) stack the PDF arrays from each scale to produce a $5 \times 100$ matrix; (5) stack the matrices from each redshift to produce a $4 \times 5 \times 100$ datacube. The normalization in step (2) serves to standardize the data so that values across different redshifts can be meaningfully compared by the network. Although the natural domain for each PDF is scale-dependent, we have chosen a binning range wide enough that the full information at each scale is included.

\begin{figure}
	\includegraphics[width=\columnwidth]{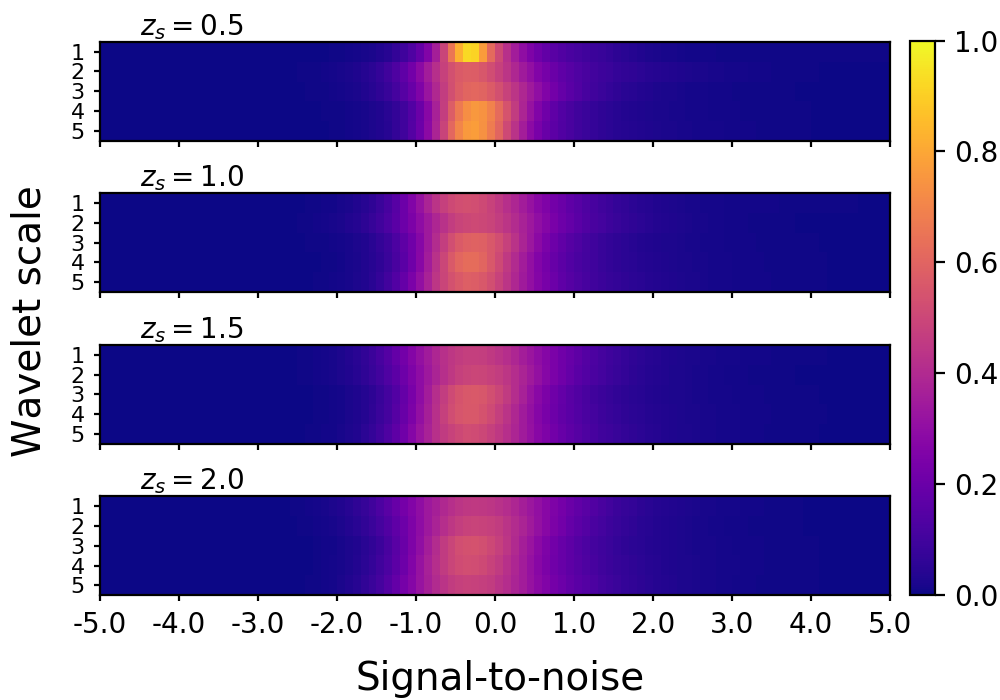}
    \caption{Reduced representation of a $\Lambda$CDM convergence map based on the PDF of wavelet coefficients at different scales and redshifts. The four matrices stack along the redshift axis to produce $4 \times 5 \times 100$ datacubes, condensed from $4\times2048^2$ pixels, which serve as inputs to the CNN.}
    \label{fig:pdf_ribbons}
\end{figure}

The resulting PDF representation of an example $\Lambda$CDM map is shown in Fig.~\ref{fig:pdf_ribbons}, separated along the redshift axis. The matrices amount to the combined PDFs across all wavelet scales of interest and reduce the input data dimensionality by a factor of about 8400. Furthermore, they are concatenated in a way that should allow the CNN to learn patterns reflecting the variation across scales as well as the evolution with redshift. We recall that wavelet filtering is not the same as Gaussian smoothing, which is a form of low-pass filtering. As our wavelet functions act as band-pass filters, they isolate features in the original map only near the scale of interest. Moreover, the wavelet transform possesses an exact inverse, meaning that information is not lost in the change of basis.

We have also explored an alternative data representation based on wavelet peak counts. Since it gives similar final results in both the clean and noisy cases, we only present the PDF representation here. Futher details are presented in Appendix \ref{sec:peaks}.

\subsection{CNN architecture}\label{subsec:architecture}
We build a CNN to perform cosmological model classification based on the $4 \times 5 \times 100$ inputs computed from each convergence map. The architecture is detailed in Table~\ref{table:arch}. All activation functions are taken to be rectified linear units \cite[i.e. ReLU,][]{ReLU.2010}. The first two convolution layers allow the network to extract features related to local variations in the datacube. Each convolution layer has eight output filters to store the different properties found, and the $2 \times 3 \times 10$ size is to explore correlations in the PDFs both within a scale as well as across neighboring scales and redshifts.

\begin{table}
\caption{Achitecture of our CNN.}
\label{table:arch}
\centering
\setlength{\tabcolsep}{0.8em}
\begin{tabular}{c c c}
\hline\hline
Layer type & Output shape & \# params \\
\hline\hline
   Input layer                         & $1 \times 4 \times 5 \times 100$ & 0 \\
   Conv 3D [$2 \times 3 \times 10$]    & $8 \times 4 \times 5 \times 100$ & 448 \\
   Conv 3D [$2 \times 3 \times 10$]    & $8 \times 4 \times 5 \times 100$ & 3848 \\
   Max pooling [$1 \times 1 \times 5$] & $8 \times 4 \times 5 \times 20$  & 0 \\
   Conv 3D [$2 \times 3 \times 10$]    & $8 \times 4 \times 5 \times 20$  & 3848 \\
   Max pooling [$1 \times 1 \times 2$] & $8 \times 4 \times 5 \times 10$  & 0 \\ 
   Dropout [$0.3$]                     & $8 \times 4 \times 5 \times 10$  & 0 \\
   Flatten                             & $1600$                           & 0 \\ 
   Fully connected                     & $32$                             & 51232 \\
   Fully connected                     & $16$                             & 528 \\
   Fully connected                     & $4$                              & 68 \\ 
\hline
\end{tabular}
\end{table}

A max-pooling operation follows the convolutions and reduces the number of parameters with a mask of size $1 \times 1 \times 5$. This condenses local features into their most relevant pixels while preserving the spatial structure of the data. Another pair of convolution and max-pooling layers constitute the fourth and fifth layers. The same kernel as above is used for the convolution, but the max-pooling layer uses a smaller mask ($1 \times 1 \times 2$) to reduce the information now by a factor of two.

Before flattening the result into a one-dimensional array of size 1600, we perform a $30\%$ dropout, which helps prevent over-fitting of the training data. The last three layers are fully connected and end with a softmax function for the actual classification. The final output layer has four nodes corresponding to the four cosmological models that label our data.

\subsection{Generating noisy data}\label{subsec:noise}
Convergence maps derived from real data (i.e. galaxy shape catalogs) are subject to many sources of noise, such as intrinsic galaxy shapes, masking and border effects during shear inversion, instrumental noise, and shape measurement errors. To assess the robustness of our network to noise, we generate noisy versions of each convergence map. Following \citetalias{PPG.etal.2018} (cf. the Appendix), we add a random Gaussian noise component with standard deviation $\sigma=0.7$ and zero mean---a conservative (pessimistic) scenario. For comparison, we also explore a more optimistic scenario with $\sigma=0.35$, which could arise, for example, from a deeper galaxy survey with a larger number density of objects.

Noise is applied directly to the convergence maps before performing the wavelet transform in the dimentionality reduction scheme. In the noisy analysis, the PDF representations are computed the same way as before, meaning we do not include any additional de-noising step. The result for each map is a datacube of the same dimension as depicted in Fig.~\ref{fig:pdf_ribbons} but now blurred.

\subsection{Training strategy}\label{subsec:training}
Our cosmological model discrimination problem in the neural network framework becomes a probabilistic classification problem; we thus use the categorical cross-entropy as the CNN's loss function to minimize. To train the network, we use the Adam \cite{Adam.2014} optimizer with $\beta_1 = 0.9$, $\beta_2 = 0.999$ (default \texttt{Keras}\footnote{\url{https://keras.io}} parameters), and an initial learning rate of $\eta_0=0.001$. We add a decay parameter $\gamma=0.001$ to mildly decrease the value of the learning rate throughout epochs according to $\eta(t)=\eta_0 (1 + \gamma\,t)^{-1}$, where $t$ identifies the update step. One step corresponds to one update through a batch of chosen size, which we take to be 200.

We train the CNN on correctly labeled input data for 1000 epochs using 75\% of each model's full dataset, randomly selected. The remaining 25\% is reserved for testing the network. Consequently, it takes four update steps to complete one epoch and 4000 steps to finish training. The learning rate is five times smaller at the end of training than at the beginning, which allows for smoother convergence towards the end. We record the loss function and validation accuracy after each epoch (see Sec.~\ref{sec:performance}), computed on the training and test data, respectively, to monitor the training progress.

\begin{figure}
	\includegraphics[width=\columnwidth]{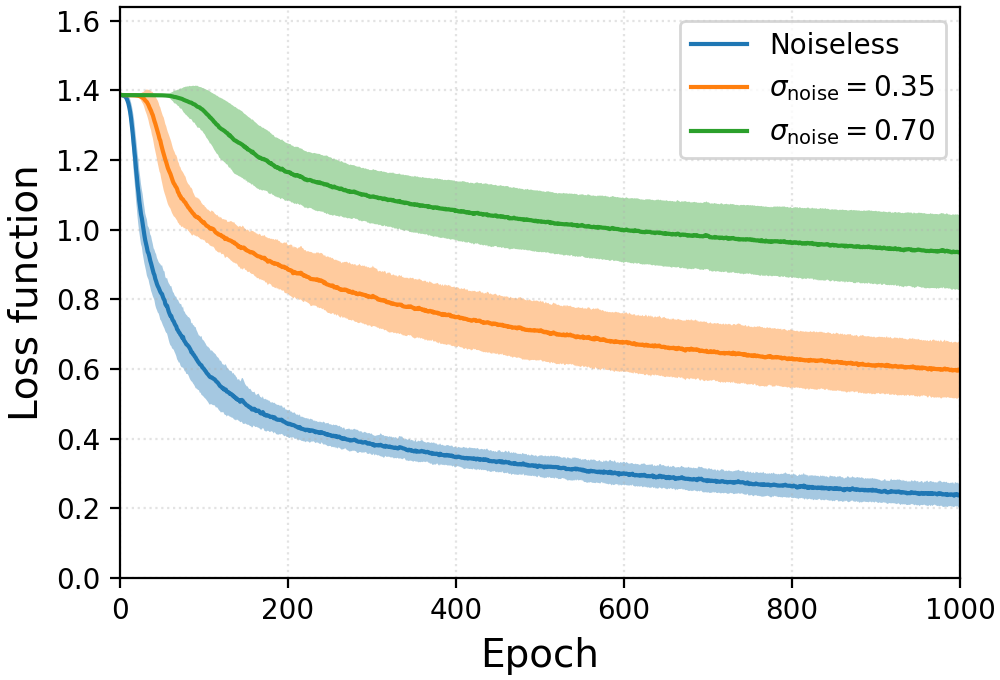}
    \caption{Evolution of the loss function with training epoch. Solid lines represent the mean over 100 full training iterations with standard errors (one sigma) shown as shaded bands. Three different noise levels are indicated by color. The value of the loss function is linked to the quality of the data: noisier convergence maps correspond to a larger asymptotic value of the loss function.}
    \label{fig:loss_function}
\end{figure}

\begin{figure}
	\includegraphics[width=\columnwidth]{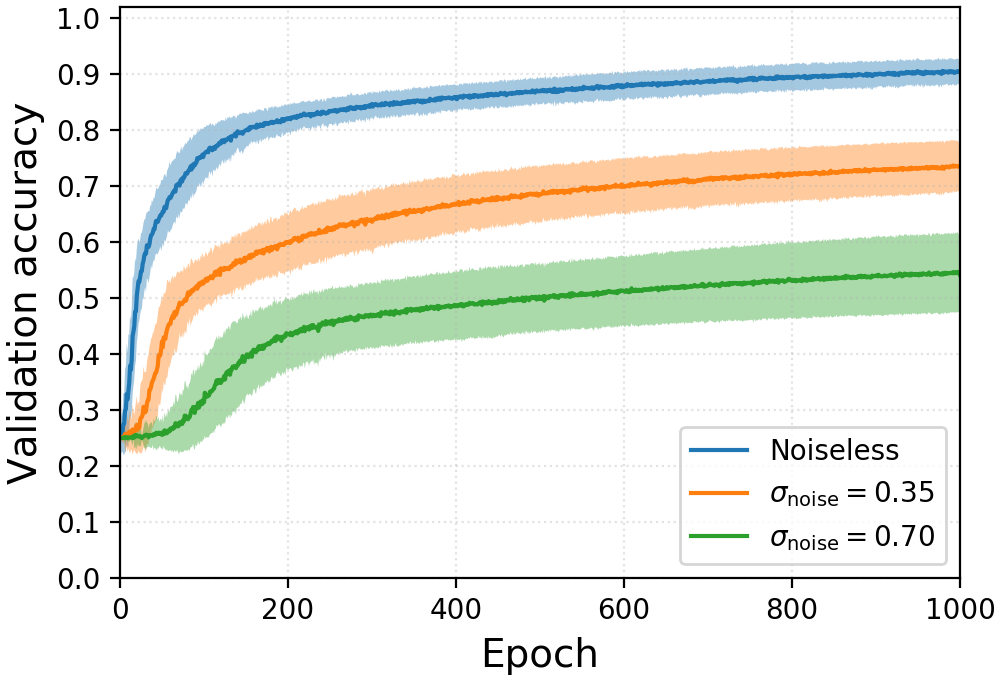}
    \caption{Evolution of the validation accuracy with training epoch. As in Fig.~\ref{fig:loss_function}, solid lines represent the mean over 100 iterations, and the shaded areas are the standard errors. Accuracy approaches a constant for each of the three noise cases, and as expected, the network provides more accurate predictions as the noise level decreases.}
    \label{fig:validation_accuracy}
\end{figure}

\section{Network performance measures}\label{sec:performance}
\subsection{Loss function and validation accuracy}\label{subsec:loss_and_accuracy}
Training can be monitored by the computation of several metrics. One is the loss (or cost) function, which measures the discrepancy between the targeted output and the prediction of the network. As mentioned in Sec.~\ref{subsec:training}, the standard loss function for classification problems is based on the categorical cross-entropy. Given two discrete probability distributions $p$ and $q$ over the same set of events $\Omega$, cross-entropy $l$ is computed as 
\begin{equation}
  l(p, q) = - \sum \limits_{\omega \in \Omega} p(\omega) \log q(\omega),
\end{equation}
where $q$ denotes the predicted probability and $p$ the desired output. For our purposes, the set of events corresponds to the four model labels. After each epoch, the loss function is the sum of the cross-entropies over the training data:
\begin{equation}
 \mathcal{L} = \sum \limits_{i=1}^{n} l(p^{(i)}, q^{(i)}),
\end{equation}
where $n=768$ is the number of training datacubes.

The network's learning procedure consists in changing its parameter values (i.e. weights) via stochastic gradient descent and back-propagation in order to make predictions that progressively better match the desired output (i.e. that minimize the loss function). The smaller the loss function, the better the network has learned the relationship between the training data and their corresponding labels. By tracking the decrease in the loss function over epochs, one can determine whether the network has trained sufficiently.

In Fig.~\ref{fig:loss_function} we show the loss function evolution for the three noise levels using the full three-dimensional datacubes (i.e. all redshifts) as described in Sec.~\ref{subsec:architecture}. As expected, the higher the noise level, the larger are the final value and its variance after 1000 epochs. We have studied training as well on only a single source redshift (see Sec.~\ref{sec:results}), but we do not repeat the figure, since the result is very similar.

Another measure of a network's performance is its validation accuracy, or the ratio of correct predictions to the total number of test observations. Validation accuracy is an indicator of the predictive power of the network and measures its ability to generalize to new data. Plotted in Fig.~\ref{fig:validation_accuracy} is the evolution of the validation accuracy computed on the test data over 1000 epochs. Its behavior is consistent with the loss function in that the final accuracy is lower and exhibits higher variance as the noise level increases. The ratio of correct predictions at the end of training decreases from 92\% to 48\% as the noise standard deviation increases from 0 to 0.7.

\subsection{Evaluation strategy}\label{subsec:evaluation}
The final assessment of the CNN is obtained with a confusion matrix, computed on the test data, which characterizes the ability of the network to classify each cosmological model. A confusion matrix is a square array whose rows correspond to the true model labels and whose columns represent the predictions. The values in a given row add up to unity and indicate the probability of the network to classify a map from the input model across each of the available output labels. The closer a confusion matrix is to the identity, the better is the classification power of our network.

Our final reported numbers represent the average confusion matrix calculated over the matrices obtained from 100 iterations of the complete training process, where each iteration used a different random selection of training and test maps. We ran our code using a {\small NVIDIA} Tesla K20c GPU with 5~GB memory along with an Intel Xeon Processor E5-2620 CPU with six cores at 2.00~GHz. An iteration of 1000 epochs required about $13$ minutes, so the entire training process took approximately $20$ hours.

In \citetalias{PPG.etal.2018}, the ability of different non-Gaussian statistics to distinguish between cosmological models was measured by the discrimination efficiency $\mathcal{E}$. Reported as a percentage, with 0\% being indistinguishable and 100\% being fully distinguishable, the discrimination efficiency was calculated via the False Discovery Rate (FDR) technique of \citet{BH.1995}. In this scheme, given a statistic, wavelet scale, and source redshift, one computes $\mathcal{E}$ for one model (prediction) taking another model as reference (truth). The result is a measure of the overlap between the two distributions of the given statistic and depends sensitively on their relative shapes and positions. We compare our CNN to the discrimination efficiency found for peaks in the following section.

\section{Results}\label{sec:results}
\begin{figure}
    \includegraphics[width=\columnwidth]{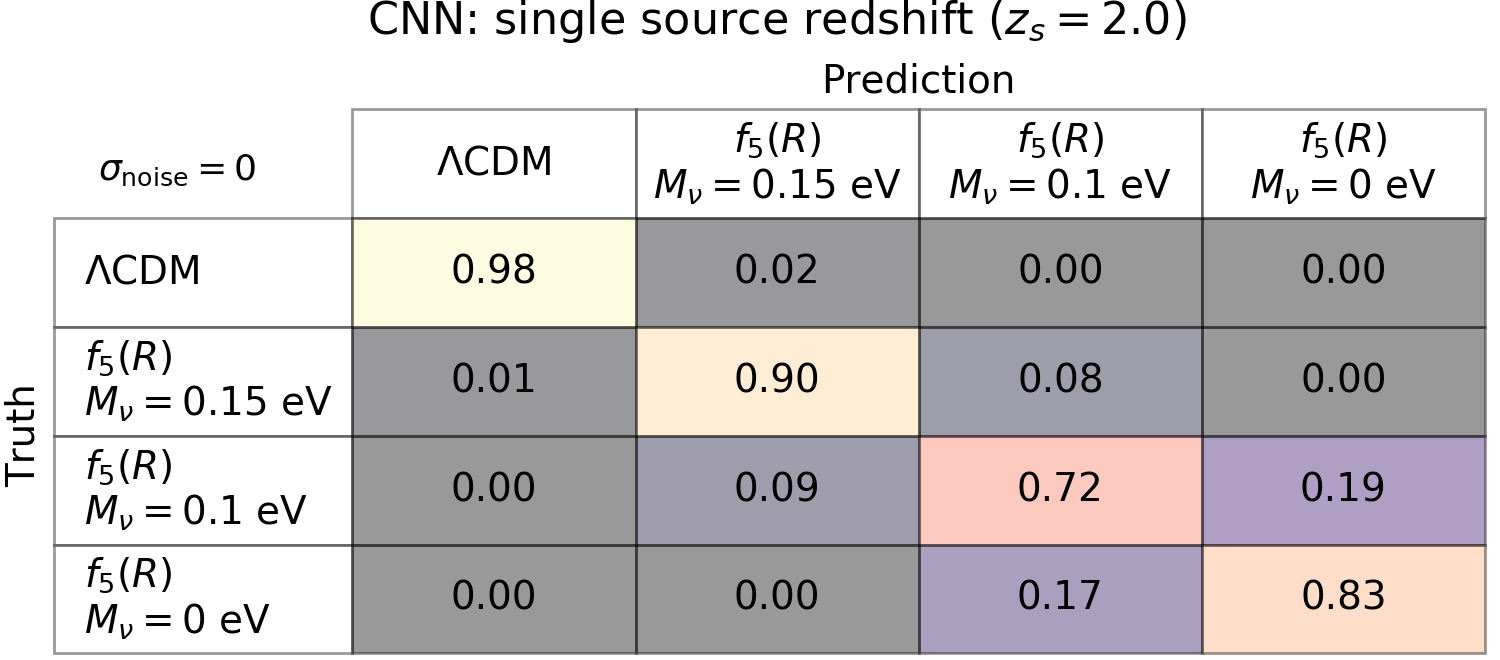}\\
    \includegraphics[width=\columnwidth]{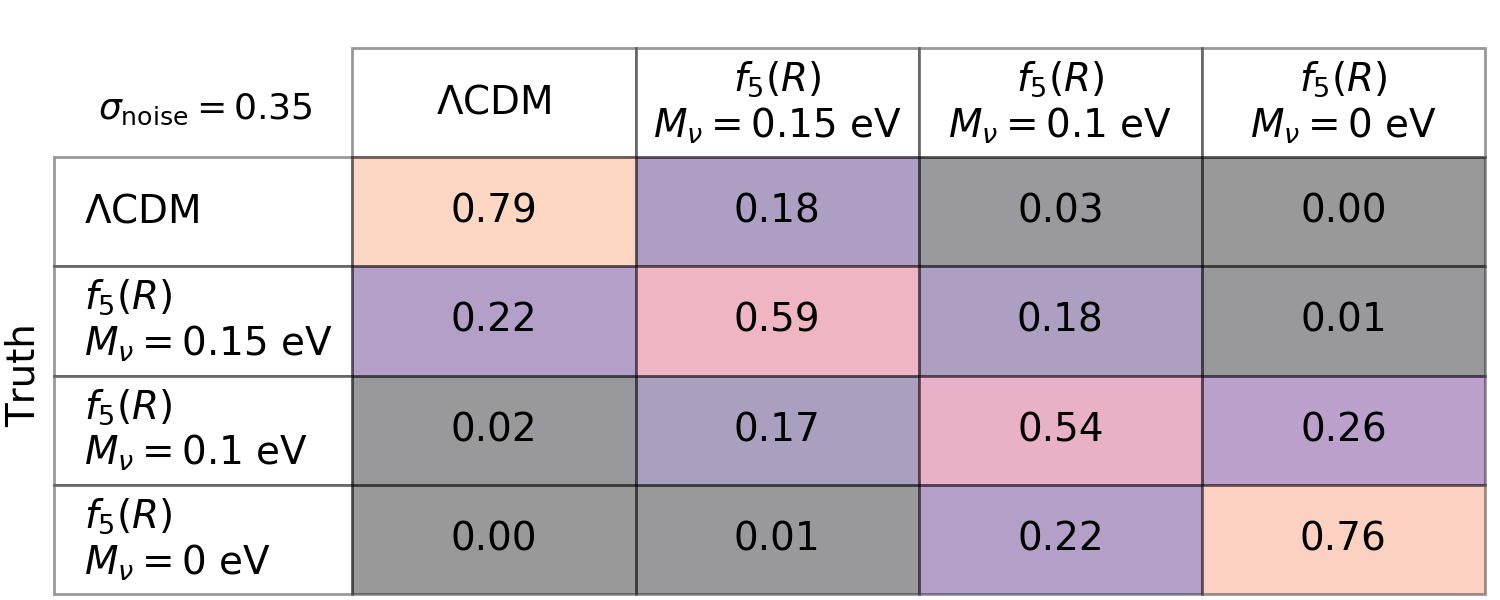}\\
    \includegraphics[width=\columnwidth]{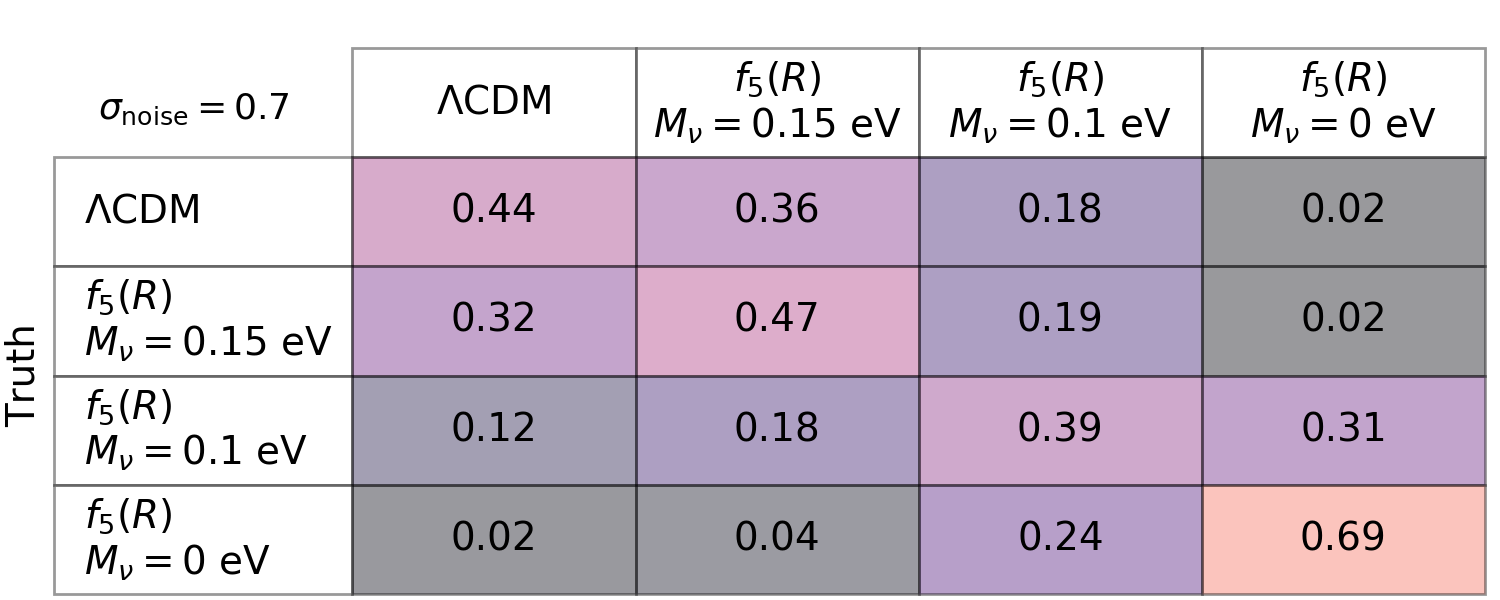}
    \caption{Confusion matrices from the trained CNN on test data using the single redshift $z_s=2.0$. Without noise, the CNN is able to discriminate all four models from each other with at least 72\% accuracy, including among the three $f(R)$ models with different neutrino masses. The rate of successful predictions decreases with increasing noise, but even for the pessimistic case ($\sigma_\mathrm{noise}=0.7$) the CNN retains non-negligible discrimination power. It is important to note that for an optimistic noise level, $\Lambda$CDM can still be distinguished with 79\% accuracy, and confusion increases over the noise-free case primarily between neighboring pairs of models.}
    \label{fig:ml_pdf_cm_zs2}
\end{figure}

As a first test of our CNN, we trained the network using only maps from the $z_s=2.0$ source redshift plane (cf. the bottom row in Fig.~\ref{fig:pdf_ribbons}). This represents a restricted case of the full network architecture in which the convolution layers are now of dimension $1 \times 3 \times 10$, and the number of trainable parameters is reduced by a factor of $\sim\!3.4$. We do this in order to be able to compare results directly to those of \citepalias{PPG.etal.2018}, since the analysis there did not accommodate multiple simultaneous redshifts. Confusion matrices are shown in Fig.~\ref{fig:ml_pdf_cm_zs2} for the three noise cases studied.

As mentioned in Sec.~\ref{sec:intro}, the $M_\nu=0.15~\mathrm{eV}$ case of $f_5(R)$ produces observations most resembling the standard model in terms of both Gaussian and non-Gaussian statistics. For example, its convergence power spectrum deviates less than 5\% from $\Lambda$CDM over the range $10^3 < \ell < 10^4$ (cf. Fig.~4 of \citetalias{PPG.etal.2018}). An $f_5(R)$ cosmology with $M_\nu=0~\mathrm{eV}$, on the other hand, would differ from $\Lambda$CDM at $>10$\% for all $\ell$ values within the same range on the basis of only Gaussian statistics. It is therefore important to check whether our neural network efficiently disentangles, in particular, the MG model ($M_\nu=0.15~\mathrm{eV}$) most degenerate with $\Lambda$CDM. This is indeed the case, as we illustrate below.

In the noise-free case (upper plot), the CNN is able to discriminate between (i.e., correctly classify) all four models with at least 72\% accuracy, notably reaching 98\% for $\Lambda$CDM and 90\% for $f_5(R)$ with $M_\nu=0.15~\mathrm{eV}$. The highest confusion rates occur among the three $f_5(R)$ models, where the closer the $M_\nu$ value is between two models, the more likely they are to be mistaken for each other. Across all four models, the $M_\nu=0~\mathrm{eV}$ and $0.1~\mathrm{eV}$ $f_5(R)$ cases are the least distinguished with a confusion rate of 17--19\%.

With noise added (middle and lower plots), confusion increases between many pairs of models, as indicated by the larger off-diagonal values. When $\sigma_\mathrm{noise}=0.35$, $\Lambda$CDM is more likely to be confused with $f_5(R)$ with $M_\nu=0.15~\mathrm{eV}$ than before, although confusion with the other $f_5(R)$ models remains low. The confusion between the $0~\mathrm{eV}$ and $0.1~\mathrm{eV}$ $f_5(R)$ cases increases as well, and these trends are further amplified at the highest noise level of $\sigma_\mathrm{noise}=0.7$.

We compare the CNN results of this work with peak counts from \citetalias{PPG.etal.2018} by translating the latter's discrimination efficiency $\mathcal{E}$ values (see Sec.~\ref{subsec:evaluation}) into confusion values as follows. Given a true reference model, we compute $\mathcal{E}$ for peak counts at $z_s=2.0$ across the same five wavelet scales used by the CNN for each test model. We have fixed $z_s=2.0$, because higher source redshifts tended to yield better results, as the convergence maps incorporate more cosmic structure information that is useful in identifying them. We record in the confusion matrix the highest value attained at any wavelet scale and then normalize each row to sum to one. The values therefore constitute the best case achievable by peak count statistics using the previous technique.

\begin{figure}
    \includegraphics[width=\columnwidth]{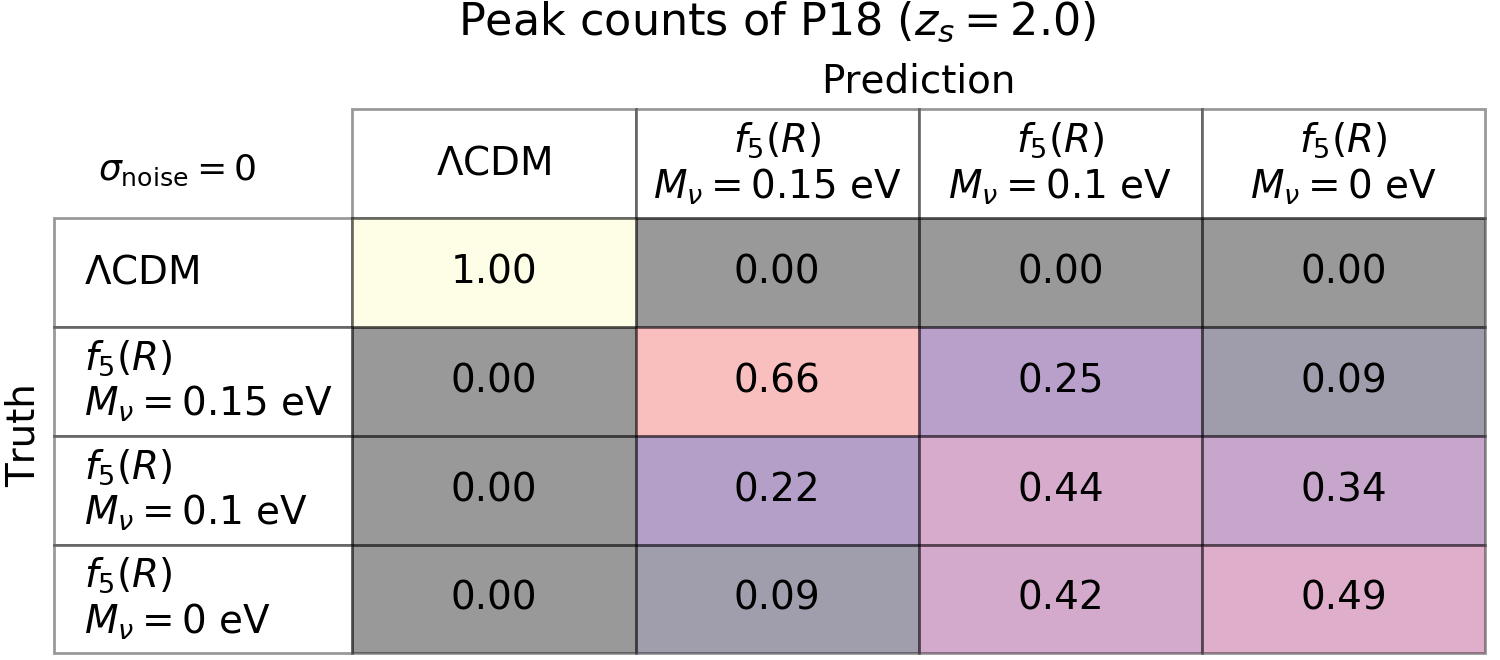}\\
    \includegraphics[width=\columnwidth]{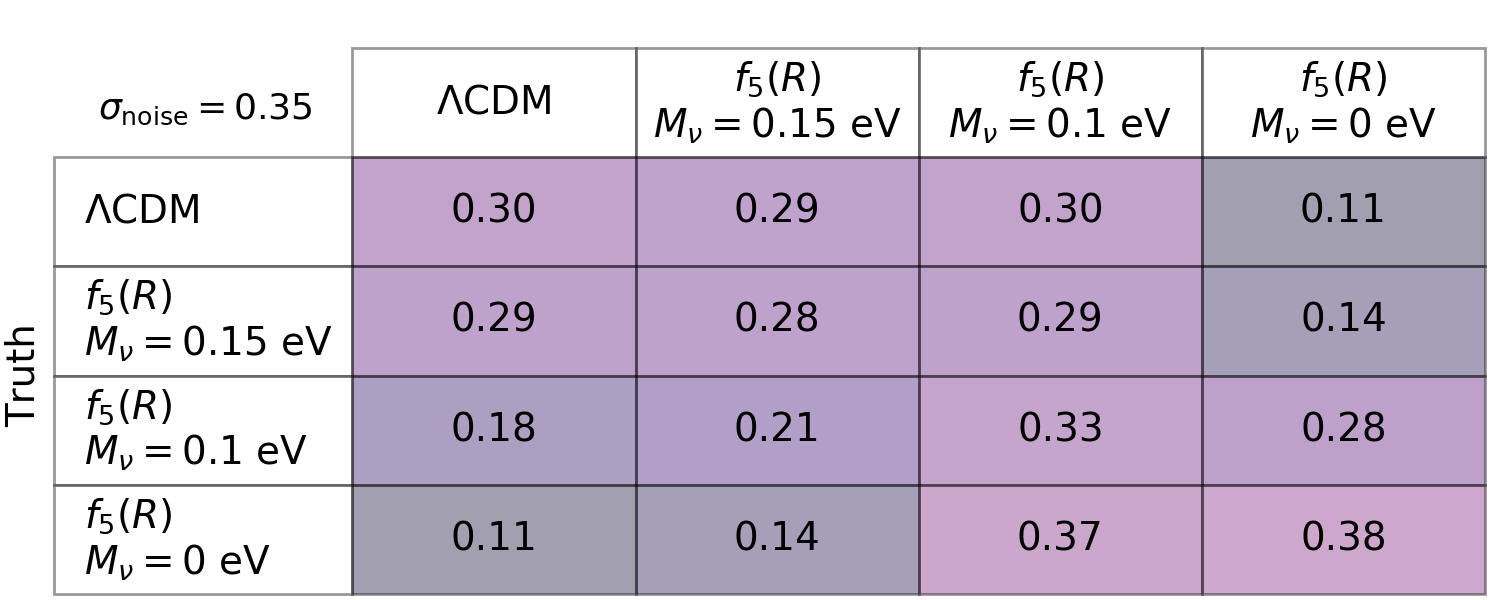}\\
    \includegraphics[width=\columnwidth]{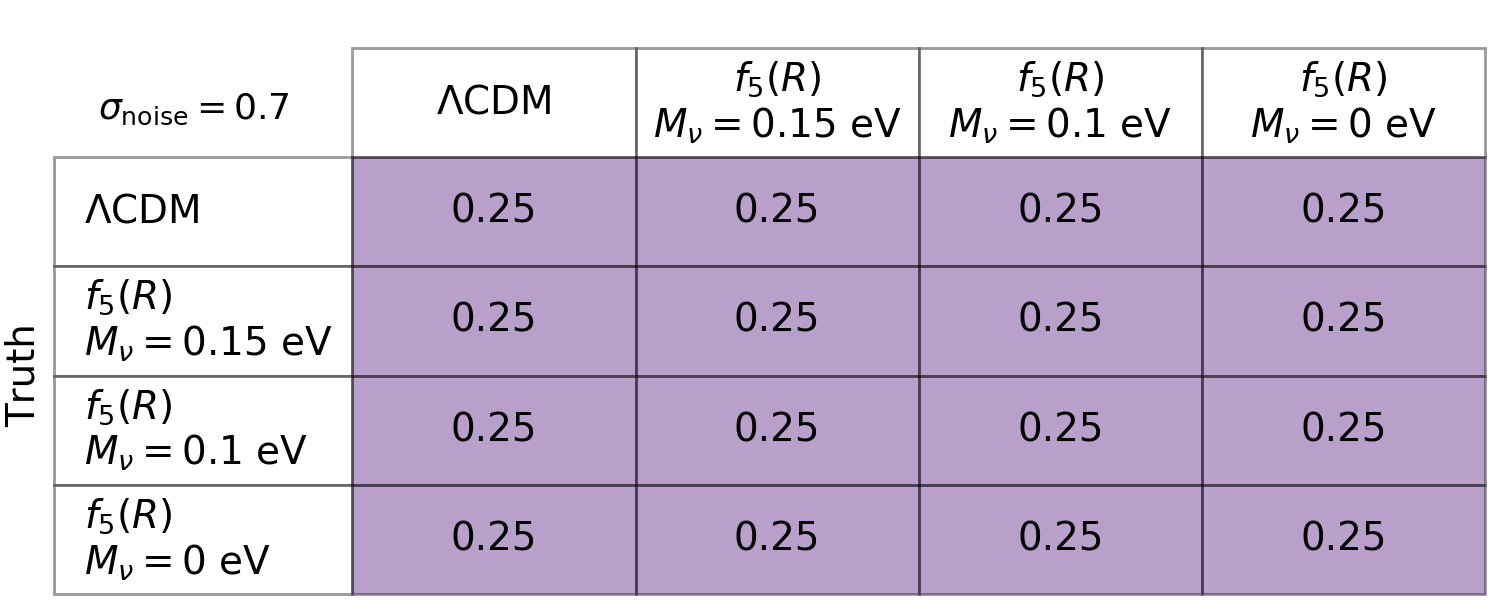}
    \caption{Confusion matrices representing the best case from thresholded peak count statistics of \citetalias{PPG.etal.2018} for $z_s=2.0$. The CNN performs significantly better in every case, i.e. it is closer to the identity matrix. Unlike the CNN, the predictive power of peaks approaches zero with increasing noise.}
    \label{fig:p18_cm}
\end{figure}

\begin{figure}
    \includegraphics[width=\columnwidth]{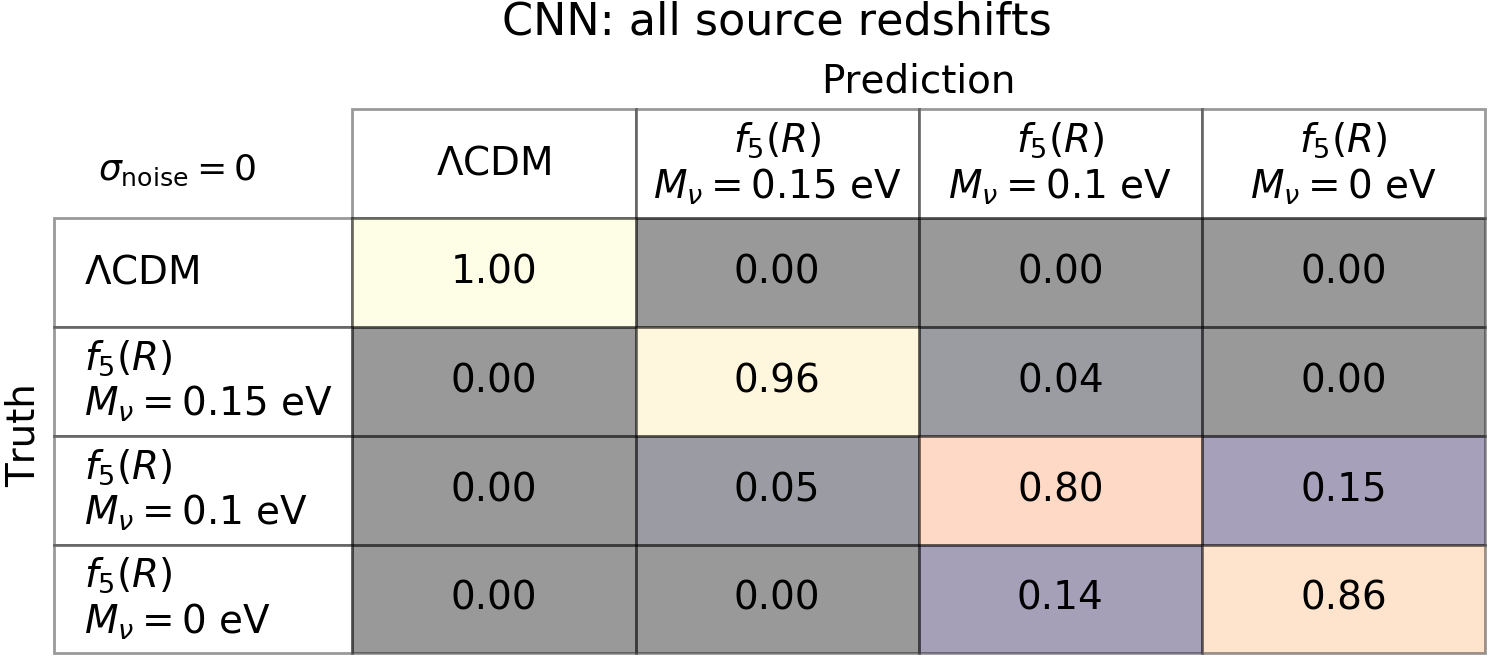}\\
    \includegraphics[width=\columnwidth]{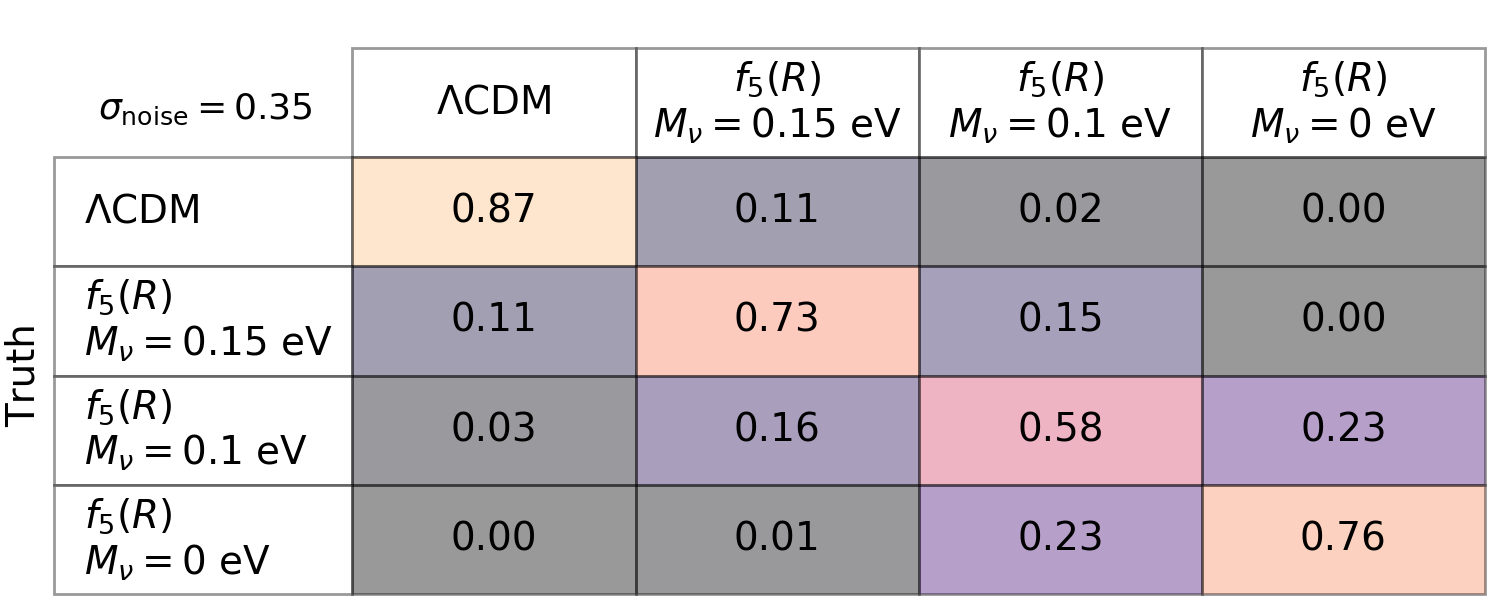}\\
    \includegraphics[width=\columnwidth]{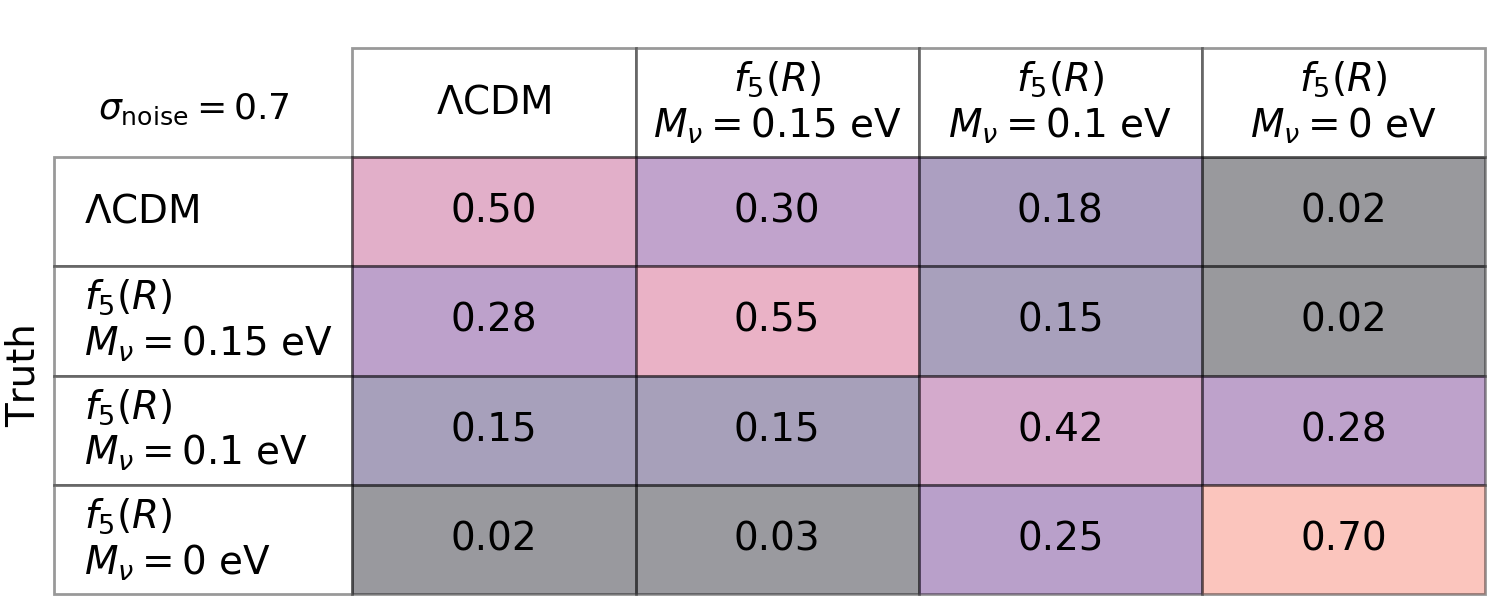}
    \caption{Confusion matrices from the trained CNN on test data using all four redshifts up to $z_s=2.0$. Plots are analogous to those in Fig.~\ref{fig:ml_pdf_cm_zs2}. The correct classification rate of the network is improved at each noise level over the corresponding single high-redshift case. Without noise, the network correctly identifies the model at least 80\% of the time.}
    \label{fig:ml_pdf_cm}
\end{figure}

The analogous plots to Fig.~\ref{fig:ml_pdf_cm_zs2} using peak counts are shown in Fig.~\ref{fig:p18_cm}. The CNN outperforms peak counts at each noise level in the sense that its confusion matrices are always closer to the identity. Without noise, both the CNN and peaks are able to fully distinguish $\Lambda$CDM from the three MG models. Indeed, this result for peaks was one of the main conclusions of \citetalias{PPG.etal.2018}. However, among the three $f_5(R)$ models, the CNN is more sensitive to the value of $M_\nu$, as the correct model is identified with peaks only 44\%--66\% of the time. Furthermore, whereas the CNN retains non-negligible predictive power with rising noise, the prediction rate of peaks diminishes to the point of consistency with random guessing (i.e. 25\%). Our analysis reveals that there are additional distinctive characteristics present in the convergence maps that our network architecture can exploit but that is not captured by our previous approach with peaks.

We now present confusion matrices for the CNN trained using the full redshift information in Fig.~\ref{fig:ml_pdf_cm}. As one would expect, the inclusion of the lower redshift data improves results over the single high-redshift case. The CNN can now learn the evolution of the wavelet PDFs with time, which evidently also constitutes useful characteristic information of the models. In the noiseless case, $\Lambda$CDM and $f_5(R)$ with $M_\nu=0.15~\mathrm{eV}$ are now essentially completely distinguishable, both from each other and from the other $f_5(R)$ models. Confusion overall is still highest between the other two $f_5(R)$ models at a rate of 14--15\%.

The fact that the confusion matrix does not achieve a perfect identity even with noise-free data could indicate that our choice of network architecture is suboptimal for this problem. Results might also be improved by training for more epochs, although this is not likely to offer substantial gains given that the slope of the loss function is already nearly zero after 1000 epochs (see Fig.~\ref{fig:loss_function}). Modifications to the data reduction scheme are possible as well. For example, the distribution of peaks detected as a function of wavelet scale and signal-to-noise could be used instead of the full distribution of wavelet coefficients. Alternatively, a different range of wavelet scales that include lower angular frequencies might provide different results. We investigate these latter two possibilities as appendices.

\section{Conclusions}\label{sec:conclusions}
Weak-lensing convergence maps carry significant information that, if properly extracted and leveraged, can identify their underlying cosmology. To this end, we designed a CNN to distinguish among a set of degenerate cosmological scenarios using the similar convergence maps they produce. We tested our method on four simulated models occupying positions in the joint parameter space of $f(R)$ models and massive neutrinos. The simulations have been designed to be difficult to discriminate from one another based on their two-point and higher-order statistics. Breaking such degeneracies in observables with $\Lambda$CDM represents an important step toward better understanding the fundamental nature of gravity and of cosmic acceleration.

To alleviate the computational cost of training on many large maps, we reduced the dimensionality of the problem using a novel data representation based on multi-scale wavelet PDF coefficients. Compared to peak count statistics, which have been shown to outperform other higher-order statistics \citepalias{PPG.etal.2018}, our CNN performed significantly better in terms of the confusion matrix between models for three different noise levels. While peak counts could efficiently discriminate $\Lambda$CDM from the $f_5(R)$ models, they could not reliably separate $f_5(R)$ models containing different neutrino masses. Furthermore, peak statistics rapidly lost all discrimination power when noise was added to the maps. We have demonstrated that our CNN ameliorates both of these problems; it not only correctly classifies all four models in the noiseless case with better than $80\%$ accuracy (using the full redshift information), but it retains non-negligible discrimination power even in the presence of noise.

Our results suggest further opportunities to study cosmology with machine learning and weak lensing. The network architecture we have built is a classification scheme to test four models in which certain pairs are particularly degenerate when looking at only Gaussian information \citepalias{PPG.etal.2018}. The CNN is indeed able to discriminate such degenerate cases, and this motivates further investigation into whether such a methodology may be extended. For example, one could in principle run simulations at many points in a densely sampled space of degenerate parameters and then train a classifier, as we have done, on their lensing maps. Assuming successful training, this could facilitate distinguishing many more classes of physically interesting degenerate cosmologies. The high computational cost of $N$-body simulations, however, makes this prospect currently difficult.

In addition to classifying among a set of simulated scenarios, in practical applications one may also want to solve a different, complimentary problem: inferring cosmological parameters from real survey data given a single model. In this case, the architecture would be modified to solve a regression instead of a classification problem, where the output of the network would be the (continuous) probability of the cosmological parameters themselves. In its current form, our CNN is not suitable for this type of analysis, as it really answers a different question. It provides rather a first proof of concept that this methodology may be used to discriminate cosmologies which are otherwise degenerate, and it further demonstrates the importance of being able to extract non-Gaussian information from data. In any case, many more simulations at other points of parameter space would still be necessary to train a regression network.

We have shown in this work that a neural network can indeed learn the necessary relationships between weak-lensing data and the distinct physical models that produce them. Moreover, incorporating prior domain knowledge into the choice of data representation can accelerate the training process while still achieving good results. We have demonstrated this using a simple solution based only on the PDF of wavelet coefficients and without the need, for example, to define what a peak is. Other dimensionality-reduction techniques may also be explored to further optimize the data compression or limit losses in local spatial correlations that could be useful for a network to discriminate models.

The Python code with instructions for reproducing our results will be available at \url{http://www.cosmostat.org/software/mgcnn}.

\textit{Acknowledgments}---AP acknowledges support from an Enhanced Eurotalents Fellowship, a Marie Sk\l{}odowska-Curie Actions Programme co-funded by the European Commission and Commissariat {\`a} l'{\'e}nergie atomique et aux {\'e}nergies alternatives (CEA). JM has received funding from the European Union's Horizon 2020 research and innovation programme under the Marie Sk\l{}odowska-Curie grant agreement No. 664931. CG and MB acknowledge support from the Italian Ministry for Education, University and Research (MIUR) through the SIR individual grant SIMCODE (project number RBSI14P4IH). JM, CG, and MM acknowledge support from the Italian Ministry of Foreign Affairs and International Cooperation (MAECI), Directorate General for Country Promotion (project ``Crack the lens"). We also acknowledge the support from the grant MIUR PRIN 2015 ``Cosmology and Fundamental Physics: illuminating the Dark Universe with Euclid"; and the financial contribution from the agreement ASI n.I/023/12/0 ``Attivit{\`a} relative alla fase B2/C per la missione Euclid". The {\small DUSTGRAIN}-{\em pathfinder} simulations discussed in this work have been performed and analyzed on the Marconi supercomputing machine at Cineca thanks to the PRACE project SIMCODE1 (grant No. 2016153604) and on the computing facilities of the Computational Center for Particle and Astrophysics (C2PAP) and of the Leibniz Supercomputer Center (LRZ) under the project ID pr94ji.

\appendix
\section{Peak counts PDF representation}\label{sec:peaks}
The wavelet PDF representation scheme presented in Sec.~\ref{sec:methodology} captures both Gaussian and non-Gaussian features of convergence maps as a function of scale. Counts of peaks, or local maxima, are more restrictively a non-Gaussian statistic used to similarly characterize an image. It has been shown (e.g. see \citep{DH.2010,KHM.2010,YKW.etal.2011,MSH.etal.2012,LK.2015b,MBK.etal.2015,LLZ.etal.2016,DES.SV.2016,PLL.etal.2017,SLH.etal.2018,MSH.etal.2018,FKS.etal.2018,LLZM.etal.2018}) that such peaks counted in weak-lensing maps as a function of signal-to-noise can be used to constrain cosmological parameters. Given the results in Sec.~\ref{sec:results} using the full PDF of wavelet coefficients, one may wonder how much of the classification power is coming from just the peaks contained in the representation.

\begin{figure}
	\includegraphics[width=\columnwidth]{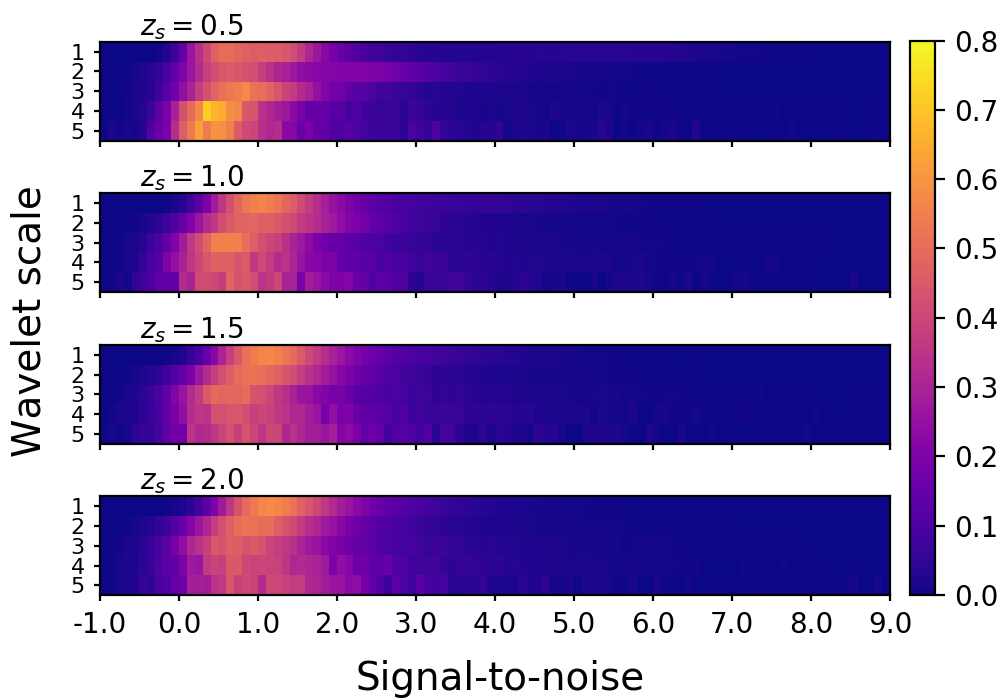}
    \caption{Condensed representation of a $\Lambda$CDM convergence map based on its multi-scale peak count distributions. As in Fig. \ref{fig:pdf_ribbons}, the four matrices stack along the redshift axis to produce $4 \times 5 \times 100$ datacubes, reduced from the original $4\times2048^2$ pixels.}
    \label{fig:pc_ribbons}
\end{figure}

\begin{figure}
    \includegraphics[width=\columnwidth]{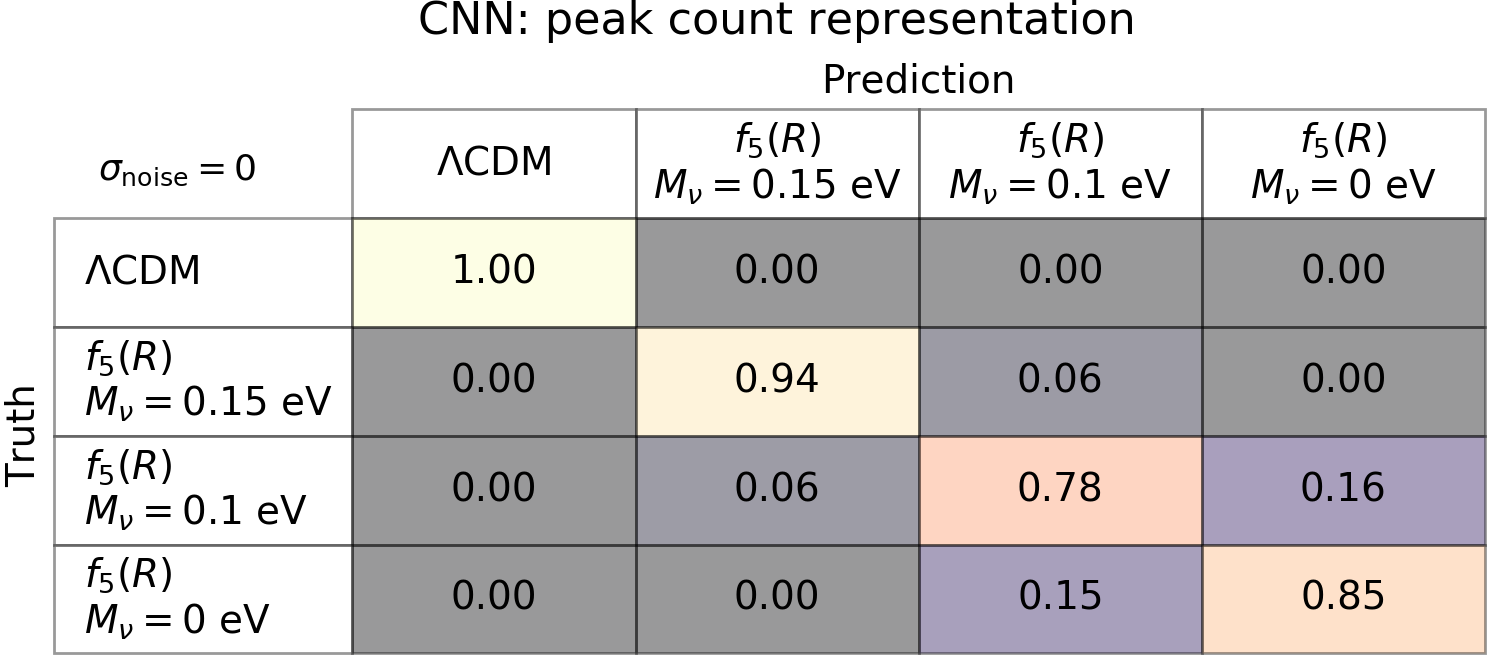}\\
    \includegraphics[width=\columnwidth]{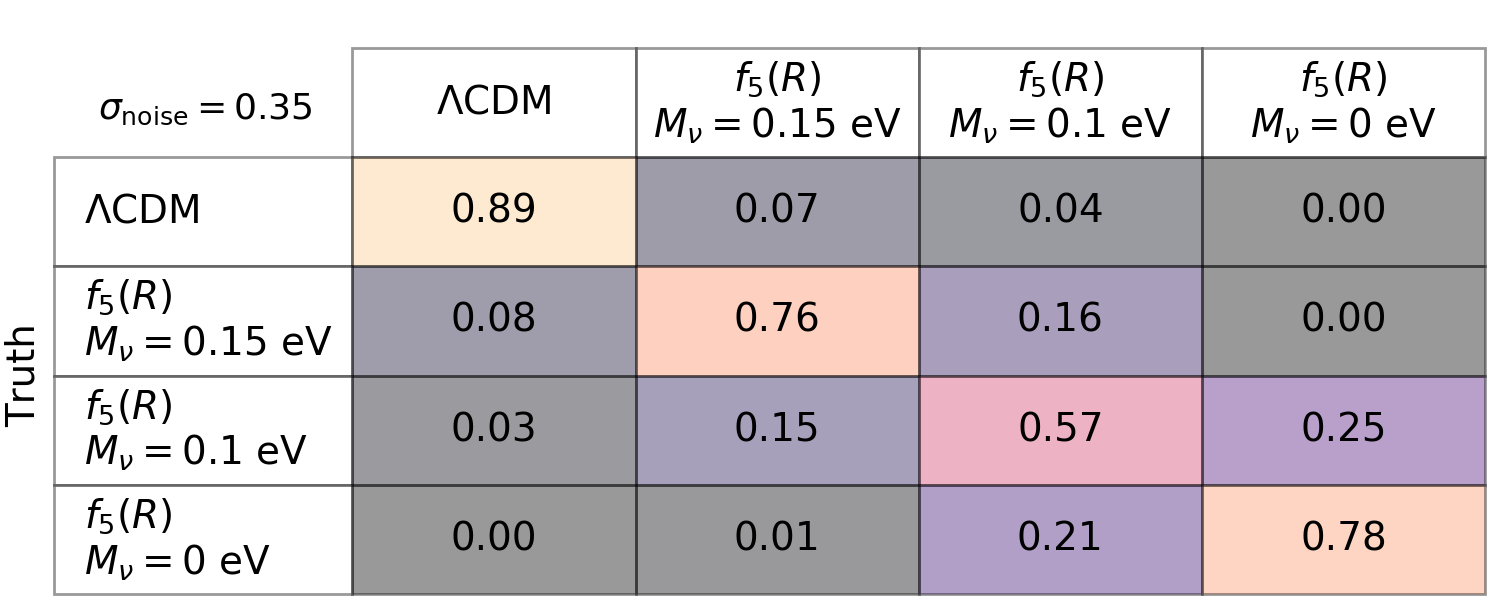}\\
    \includegraphics[width=\columnwidth]{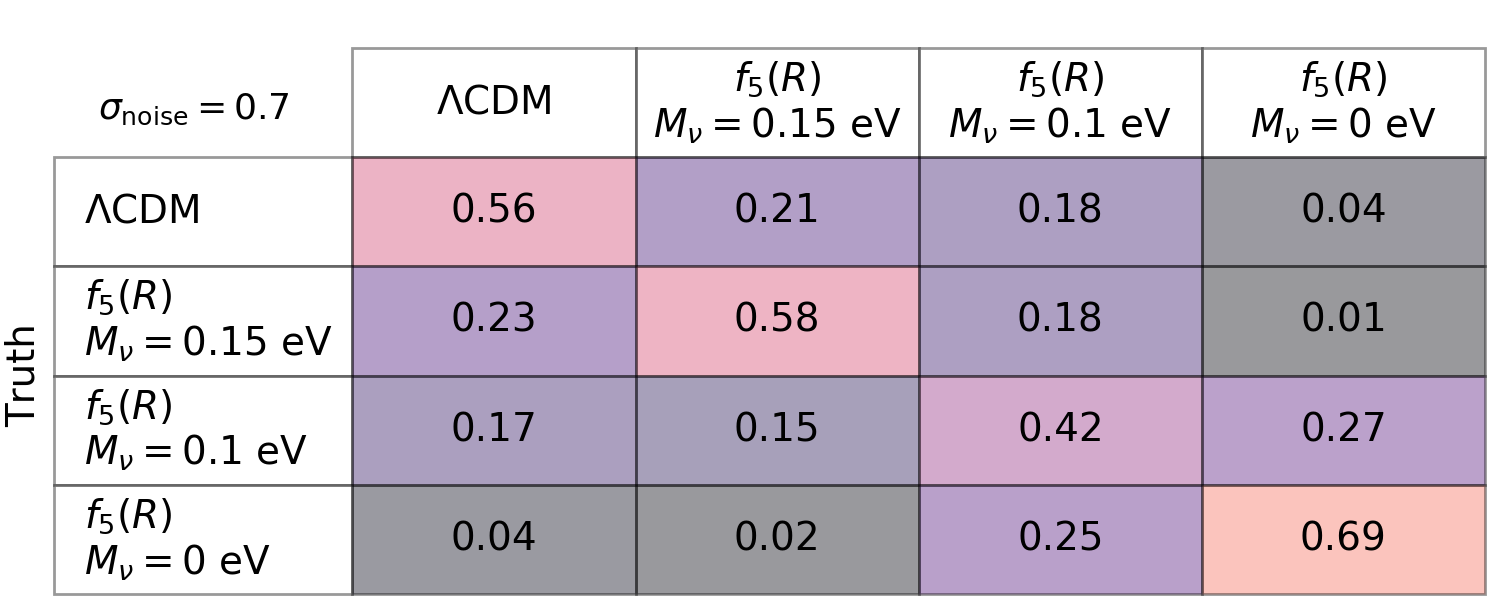}
    \caption{Confusion matrices using the peak count PDF representation to be compared with Fig.~\ref{fig:ml_pdf_cm}. The network trained on peaks alone (i.e., the non-Gaussian part of the signal) achieves comparable classification power to the full wavelet PDFs.}
    \label{fig:ml_pc_cm}
\end{figure}

We explore in this appendix an alternative dimensionality-reducing representation of convergence maps based on peak counts alone. Peaks are counted as a function of wavelet scale in a similar process to that described for the PDFs in Sec.~\ref{subsec:wavelets}: (1) compute the starlet transform with $j_{\max}=5$; (2) divide each resulting wavelet map $w_j$ by the standard deviation of all 1024 maps (256 per model) at the same redshift and scale; (3) identify peaks as pixels whose amplitudes are larger than their eight nearest neighbors; (4) bin the peak amplitudes between -1.0 and 9.0 per scale with a step size of 0.1 and normalize by the integral over the range; (5) stack the peak count distributions from each scale to produce a $5 \times 100$ matrix (6) stack the matrices from each redshift to produce a $4 \times 5 \times 100$ datacube.

An example datacube analogous to Fig.~\ref{fig:pdf_ribbons} is shown in Fig.~\ref{fig:pc_ribbons} for a $\Lambda$CDM map. As for the full PDF representations, the signal-to-noise range is chosen to be wide enough to contain the full distributions regardless of redshift, scale, and model. Peak counts in the lower bins are larger than for the higher bins as expected, and the evolution across scales with redshift can be seen as well. We note that peak signal-to-noise can be negative by our definition, because peak amplitudes themselves can be negative. Such peaks reside in underdense regions of sky but are not the same as troughs, as they are still points of higher amplitude relative to their surroundings.

Results are shown in Fig.~\ref{fig:ml_pc_cm} using all four available redshifts and so should be compared to Fig.~\ref{fig:ml_pdf_cm}. We use the same network architecture and training scheme as for the full wavelet PDFs described in Sec.~\ref{sec:methodology}. Based on their confusion matrices, we see that classification power using the two different representation schemes is nearly identical. The rate of correct classification for each model (given by the diagonal elements) differs on average by only 1.3\%, 2.0\%, and 2.5\% for $\sigma_{\mathrm{noise}}=0$, 0.35, and 0.7, respectively, between the two methods. Nor do the differences appear systematically in favor one method over the other.

\begin{figure}
    \includegraphics[width=\columnwidth]{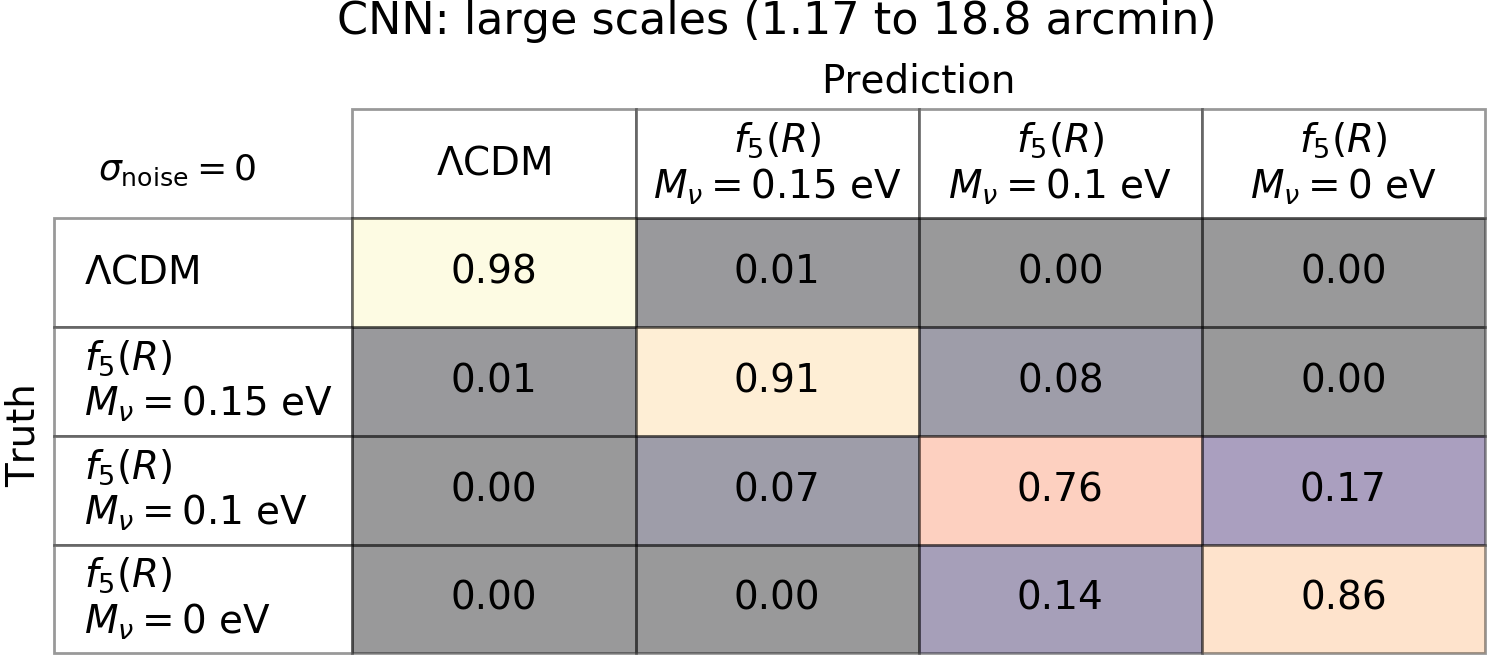}\\
    \includegraphics[width=\columnwidth]{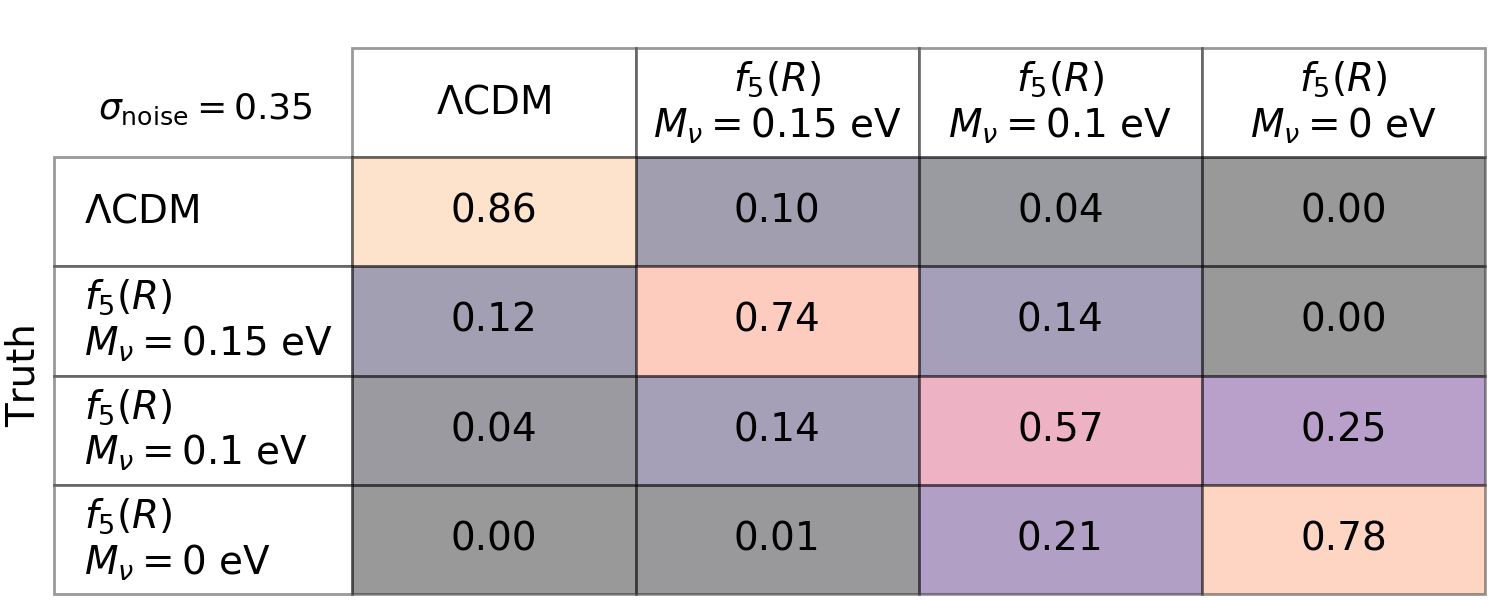}\\
    \includegraphics[width=\columnwidth]{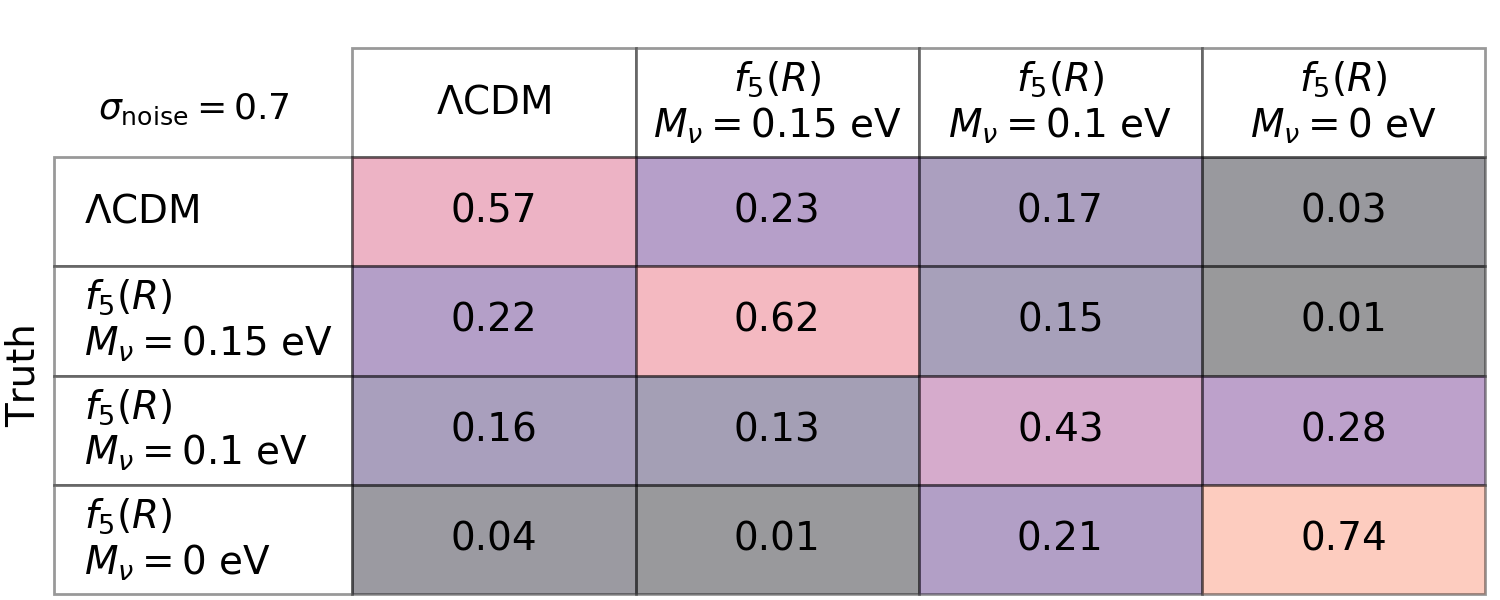}
    \caption{Results analogous to Fig.~\ref{fig:ml_pdf_cm} using the five largest wavelet scales up to $j_{\max}=7$ while excluding the two smallest. Overall classification power is again comparable to the original data representation, which uses instead the smallest five scales.}
    \label{fig:ml_pdf_cm_largescales}
\end{figure}

At least two interpretations of these results are possible. On one hand, the network could be indicating that peak counts indeed constitute the most efficient statistical lever for distinguishing between the models. Since the full wavelet PDFs necessarily contain the peak information (as well as the Gaussian), the network might simply be learning to exploit the peaks alone in both cases, thereby giving comparable results. Alternatively, the network might be learning entirely different filters depending on the input data, meaning that it still incorporates the Gaussian part of the signal in the wavelet PDFs approach. In either case, the two results point to a robustness of our architecture to the specific choice of data representation.

\section{Small vs. large scales}\label{sec:small_vs_large}
Throughout our main analysis, we have used five wavelet scales corresponding to aperture mass filter sizes of up to $\vartheta=4.69$ arcmin. The two smallest scales (0.293 and 0.586 arcmin) are below the resolution expected even for next-generation space-based surveys like \textit{Euclid}, which is aiming for weak-lensing map pixels of $\sim\!1$ arcmin. To test our network in the context of more realistic angular filtering scales, we extend the wavelet range to $j_{\max}=7$ and exclude the two smallest scales, $j=1,2$. The new range corresponds to filter apertures of 1.17, 2.34, 4.69, 9.38, and 18.8 arcmin.

Shown in Fig.~\ref{fig:ml_pdf_cm_largescales} is the classification accuracy of our network trained on clean and noisy data and using the shifted range of scales. Similar to the peak count PDF representation, the confusion matrices are not significantly different from those in Fig.~\ref{fig:ml_pdf_cm} when using the smallest scales. Average differences in the correct classification rates are now 2.8\%, 1.3\%, and 4.8\% for the three (increasing) noise cases.

Notably, the larger scales perform better than the smaller scales when the data contain the highest noise. $\Lambda$CDM and its closest $f_5(R)$ model ($M_\nu=0.15~\mathrm{eV}$) at second order (cf. Fig.~4 of \citetalias{PPG.etal.2018}), are now better distinguished in both cases by 7\% than with the smaller scales. As in Appendix \ref{sec:peaks}, the similarity of these results to those of the other data representations reveals a general capacity of our network to learn the relationship between weak-lensing information and the underlying model.

\section{Training on the power spectrum}\label{sec:power_spectrum}
To further test our data compression approach with wavelets, we perform a final experiment in which we train our CNN on the concatenated power spectrum measured across the four source redshifts. We measure the angular power spectrum $P_{\kappa}(\ell)$ on each map in 100 log-spaced bins over the angular multipole range $500 \leq \ell \leq 50000$. We choose these values to cover approximately the same angular scales as a wavelet transform with $j_{\max}=7$ (i.e., seven scales between $\ell \approx 600$ and $37000$). The information probed by the power spectrum therefore has higher resolution in angular space than the corresponding wavelets, but the resulting summary arrays are only two-dimensional: $4 \times 100$ instead of $4 \times 5 \times 100$.

\begin{figure}
    \includegraphics[width=\columnwidth]{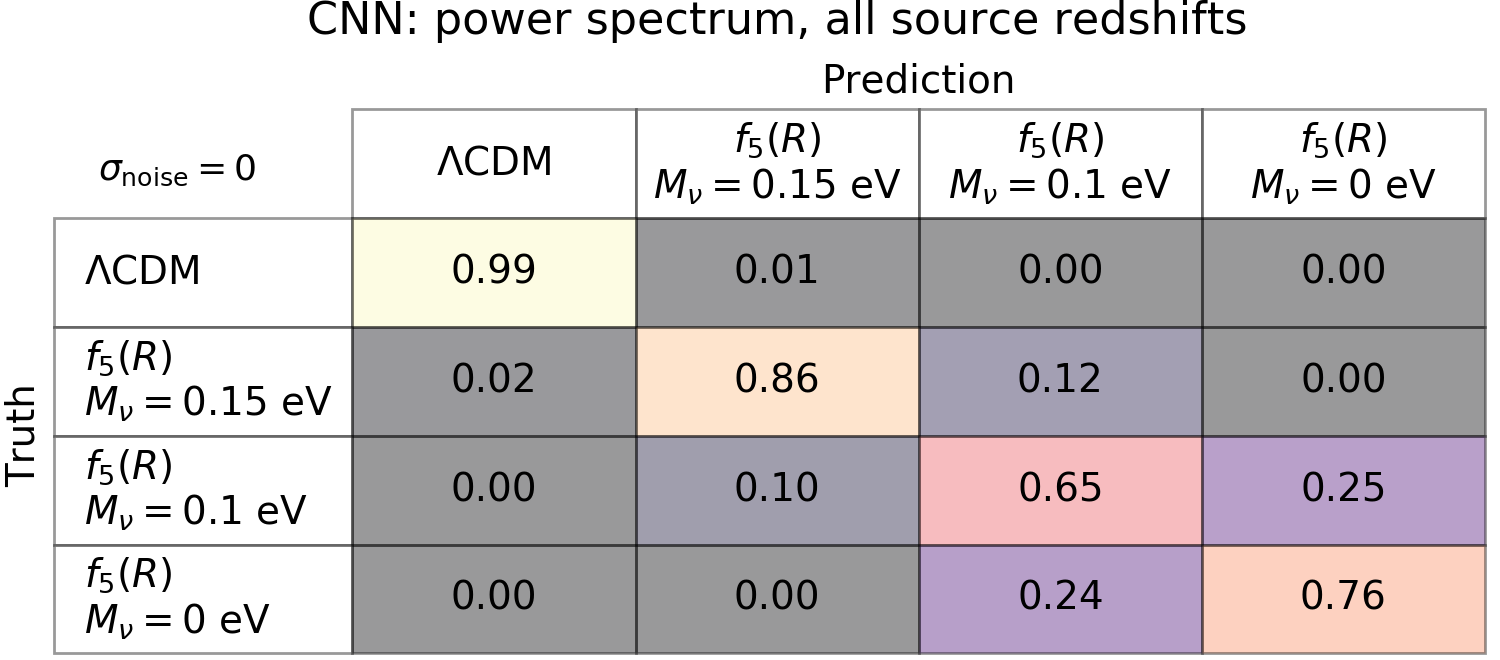}\\
    \includegraphics[width=\columnwidth]{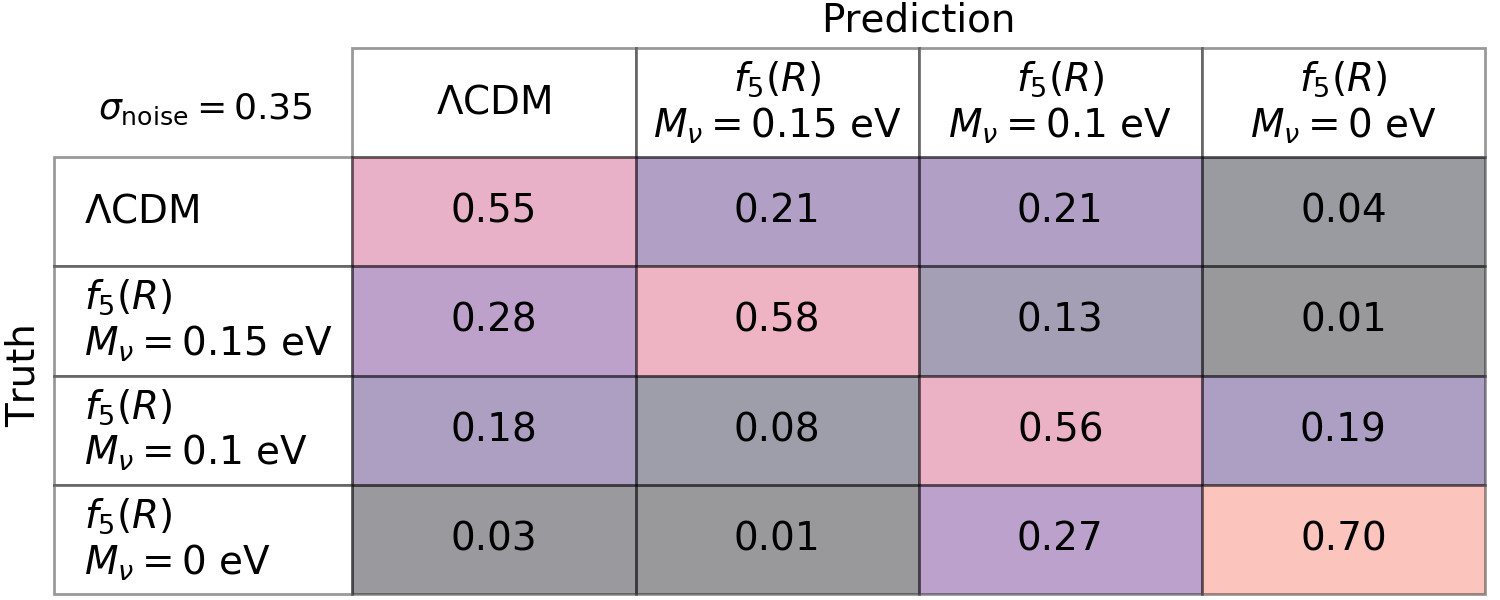}\\
    \includegraphics[width=\columnwidth]{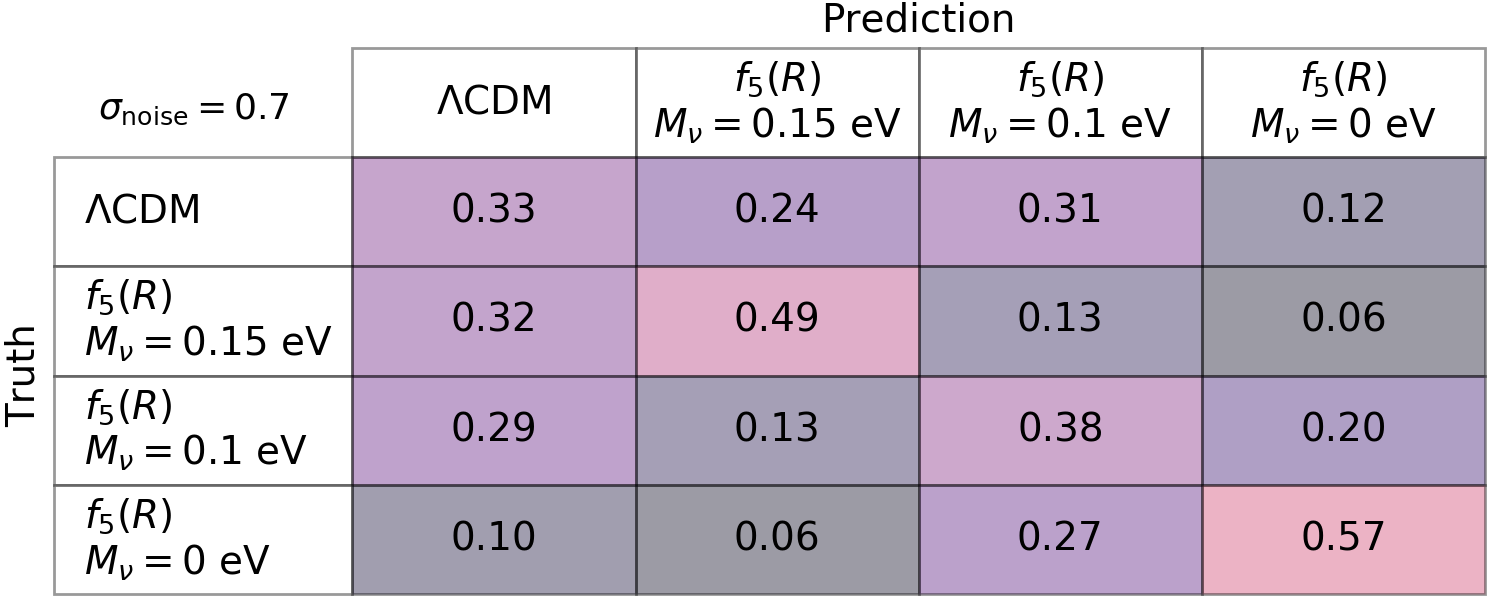}
    \caption{Confusion matrices derived from training the CNN on convergence map power spectra. A condensed representation of each line of sight was generated by concatenating the power spectrum $P_\kappa(\ell)$ measured in 100 bins between $500 \le \ell \le 50000$ from the four different source redshifts. These results should be compared to Fig.~\ref{fig:ml_pdf_cm}. The wavelet PDF representation outperforms the power spectrum at distinguishing each model, especially in the presence of noise.}
    \label{fig:cm_powspec}
\end{figure}

Using the same architecture (Table~\ref{table:arch}) and training scheme (Sec.~
\ref{subsec:training}), we derive the confusion matrices shown in Fig.~\ref{fig:cm_powspec} for the $P_\kappa$-based representation. This figure should be compared to Fig.~\ref{fig:ml_pdf_cm}, since maps from all four source redshifts on the line of sight are used in each case.

Without noise, we find that the power spectrum representation results in an overall lower discrimination power among the models compared to the wavelet PDFs. It does discriminate $\Lambda$CDM from each of the three MG models (at 99\%), but the discrimination among the different MG models is comparatively somewhat degraded. With shape noise added, the advantage of wavelets is clearly seen over the power spectrum via the middle and lower matrices. The chance of correctly identifying $\Lambda$CDM, for example, among the four models drops when using the power spectrum by 32\% and 17\% for the lower and higher noise levels, respectively. This points again to the wavelet transform providing a more useful representation space for the data, since it compresses significant signal into relatively fewer coefficients compared to in direct space.

We note that due to the power spectrum spanning many orders of magnitude, training the network on $P_\kappa(\ell)$ directly was not possible with the same architecture as used previously. The CNN could not learn to distinguish the models at all in this case; it persistently classified all maps from the training set as coming from a single (arbitrary) model, regardless of the number of epochs. To solve this, we instead trained on the logarithm of $\ell (\ell + 1) P_\kappa(\ell) / (2 \pi)$, which is both a standard way of representing the power spectrum and also served to reduce the dynamic range of the values. Training proceeded smoothly after making this change.

\bibliographystyle{apsrev4-1}
\bibliography{refs}

\end{document}